\documentclass[twocolumn,showpacs,preprintnumbers,aps]{revtex4-1}

\usepackage{graphicx}

\usepackage{dcolumn}

\usepackage{bm}

\begin{document}

\preprint{}

\title{Magnetic excitations from an $S$=1/2 diamond-shaped tetramer compound Cu$_2$PO$_4$OH}
\thanks{This manuscript has been authored by UT-Battelle, LLC under Contract No. DE-AC05-00OR22725 with the U.S. Department of Energy.  The United States Government retains and the publisher, by accepting the article for publication, acknowledges that the United States Government retains a non-exclusive, paid-up, irrevocable, world-wide license to publish or reproduce the published form of this manuscript, or allow others to do so, for United States Government purposes.  The Department of Energy will provide public access to these results of federally sponsored research in accordance with the DOE Public Access Plan(http://energy.gov/downloads/doe-public-access-plan).}

\author{M. Matsuda}

\affiliation{Quantum Condensed Matter Division, Oak Ridge National Laboratory, Oak Ridge, Tennessee 37831, USA}

\author{S. E. Dissanayake}

\affiliation{Quantum Condensed Matter Division, Oak Ridge National Laboratory, Oak Ridge, Tennessee 37831, USA}

\author{D. L. Abernathy}

\affiliation{Quantum Condensed Matter Division, Oak Ridge National Laboratory, Oak Ridge, Tennessee 37831, USA}

\author{K. Totsuka}

\affiliation{Yukawa Institute for Theoretical Physics, Kyoto University, Kitashirakawa Oiwakecho, Sakyo-ku, Kyoto 606-8502, 
Japan}

\author{A. A. Belik}

\affiliation{International Center for Materials Nanoarchitectonics (WPI-MANA), National Institute for Materials Science (NIMS), Namiki 1-1, Tsukuba, Ibaraki 305-0044, Japan}

\date{\today}

\begin{abstract}

Inelastic neutron scattering experiments have been carried out on a powder sample of Cu$_2$PO$_4$OH, which consists of diamond-shaped tetramer spin units with $S$=1/2. We have observed two nearly dispersionless magnetic excitations at $E_1$$\sim$12 and $E_2$$\sim$20 meV, whose energy width are broader than the instrumental resolution. The simplest square tetramer model with one dominant interaction, which predicts two sharp excitation peaks at $E_1$ and $E_2$(=2$E_1$), does not explain the experimental result. We found that two diagonal intratetramer interactions compete with the main interaction and weak intertetramer interactions connect the tetramers. The main intratetramer interaction is found to split into two inequivalent ones due to a structural distortion below 160 K. Cu$_2$PO$_4$OH is considered to be a good material to study the $S$=1/2 Heisenberg tetramer system.

\end{abstract}

\pacs{75.25.-j, 75.30.Kz, 75.50.Ee}

\maketitle

\section {introduction}
Spin-1/2 antiferromagnets showing a spin gap between a singlet ground state and magnetic excited states attract special interest, since the gap is a fundamental quantum effect. The spin gap behavior is widely seen in one-dimensional antiferromagnets with alternating interactions including spin-Peierls materials, in which spin-lattice coupling plays an important role. It is also seen in isolated antiferromagnetic spin clusters.~\cite{Haral05,Furrer13}

Among the various spin cluster or spin molecule systems, dimer systems have been studied intensively since there are many candidate materials. On the other hand, materials for the tetramer system are limited. The candidate materials for the linear tetramer system are Cu$_2$PO$_4$, \cite{ethredge95} NaCuAsO$_4$, \cite{ulutagay03} SrCu$_2$(PO$_4$)$_2$, \cite{Belik05} and PbCu$_2$(PO$_4$)$_2$. \cite{Belik06} Cu$_2$PO$_4$ was confirmed to be a linear-tetramer system with two intratetramer interactions, $J_1$=-55 K and $J_2$=95 K. \cite{hase97} $[$Mo$_{12}$O$_{28}$($\mu_2$-OH)$_9$($\mu_3$-OH)$_3$\{Ni(H$_2$O)$_3$\}$_4]$$\cdot$13H$_2$O consists of distorted tetrahedra of Ni$^{2+}$ moments with $S$=1. \cite{Muller00} The energy level scheme was found to be explained by two antiferromagnetic intratetramer couplings and an easy-axis anisotropy. \cite{Furrer10} An approximate square tetramer is realized in Na$_4[$V$_{12}$As$_8$O$_{40}$(H$_2$O)$]\cdot$23H$_2$O, in which V$^{4+}$ ion carries $S$=1/2. \cite{Baster02} The energy level scheme is well reproduced by two antiferromagnetic intratetramer interactions, which are isotropic.

\begin{figure}
\includegraphics[width=6.8cm]{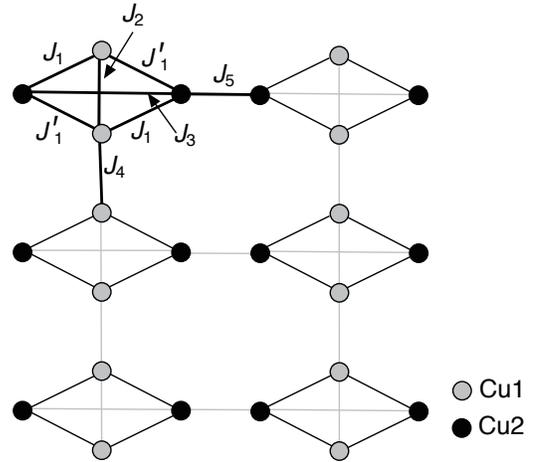}
\caption{A schematic view of the weakly interacting spin tetramers of Cu$_2$PO$_4$OH. Intratetramer interactions ($J_1$, $J'_1$, $J_2$, and $J_3$) and intertetramer interactions ($J_4$ and $J_5$) are also shown. Below the structural transition temperature at 160 K, $J_1$ and $J'_1$ become inequivalent.
}
\label{structure}
\end{figure}
Cu$_2$PO$_4$OH is a candidate material for the $S$=1/2 diamond-shaped antiferromagnetic tetramer system,~\cite{Belik07} as shown in Fig. \ref{structure}. It was estimated from the magnetic susceptibility at high temperatures that $\Theta_{CW}$ = $-$128 K and the effective moment $p\rm_{eff}$ = 1.95 $\mu\rm_B$, consistent with 1.73 $\mu\rm_B$ expected for $S$=1/2. Cu$_2$PO$_4$OH does not undergo long range magnetic ordering down to 1.8 K.
The magnetic susceptibility at low temperatures shows a spin-gap behavior and can be fitted using the isolated tetramer model with an exchange interaction of 138 K (=11.9 meV). An NMR study also confirmed the existence of the spin gap.~\cite{kuo08}
Recently, it has been found that this material shows a structural transition from being orthorhombic ($Pnnm$) to monoclinic ($P2_1/n$) around $T\rm_{st}$=160 K on cooling.~\cite{Belik11}
This structural distortion makes $J_1$ and $J'_1$ inequivalent, as shown in Fig. \ref{structure}. Although six interactions in total need to be considered in Cu$_2$PO$_4$OH, the gap behavior suggests that this material could be a tetramer system. Since neutron scattering studies in the spin tetramer system are quite limited, it is important to observe the magnetic excitations and clarify the magnetic interactions in this compound.

We have performed inelastic neutron scattering experiments on a powder sample of Cu$_2$PO$_4$OH in order to study the magnetic excitations from the tetramer spin system. We have clearly observed two magnetic excitations at $E_1\sim$12 and $E_2\sim$20 meV, whose energy widths are broader than the instrumental resolution. It was found that the energy levels cannot be explained with the simple square tetramer model with only one dominant antiferromagnetic interaction, which predicts $E_2/E_1$=2. We found that finite diagonal intratetramer interactions are needed to explain the observed ratio $E_2/E_1$$\sim$5/3. Weak intertetramer interactions, which are an origin of the broadened peak width, are considered to connect the tetramers quasi-one dimensionally. The diagonal couplings $J_2$ and $J_3$ compete with $J_1$. The difference between $J_1$ and $J_1'$ due to the structural distortion, which is another origin of the peak broadening, was estimated to be about 10\%. Although the energy levels are slightly modified because of the diagonal interactions and the splitting of $J_1$, Cu$_2$PO$_4$OH is considered to be a good material to study the $S$=1/2 Heisenberg antiferromagnetic tetramer system.

\section {Theoretical background}
The energy level diagrams for tetrahedron, rectangular, and linear tetramers in $S$=1/2 isolated Heisenberg tetramer systems are shown in Ref. \onlinecite{Haral05}. The simplest model for the diamond-shaped tetramer in Cu$_2$PO$_4$OH is the tetramer model with one dominant interaction $J_1$.
We describe here the energy level diagram in a more general model for the diamond-shaped tetramer with two diagonal interactions ($J_2$ and $J_3$), as shown in Fig. \ref{structure}. For the sake of simplicity, we assume that $J_1$=$J_1'$. The effect of $J_1'$ will be described in Sec. III.

There are two singlet, three triplet, and one quintet states in the model. Diagonalizing the single-tetramer Hamiltonian with $J_1$, $J_2$, and $J_3$, the energy levels are obtained as follows:
\begin{eqnarray*}
E_{\rm singlet\mbox{-}1}&=&\frac{1}{4}(-8J_1+J_2+J_3),\\
E_{\rm singlet\mbox{-}2}&=&-\frac{3}{4}(J_2+J_3),\\
E_{\rm triplet\mbox{-}1}&=&\frac{1}{4}(-4J_1+J_2+J_3),\\
E_{\rm triplet\mbox{-}2}&=&\frac{1}{4}(J_2-3J_3),\\
E_{\rm triplet\mbox{-}3}&=&\frac{1}{4}(-3J_2+J_3), \rm and\\
E_{\rm quintet}&=&\frac{1}{4}(4J_1+J_2+J_3).
\label{EL}
\end{eqnarray*}

The wave functions of these levels are given as
\begin{widetext}
\begin{eqnarray*}
|\rm singlet\mbox{-}1;S_{tot}^z=0>&=&\frac{|\downarrow\downarrow\uparrow\uparrow>}{2\sqrt{3}}+\frac{|\downarrow\uparrow\downarrow\uparrow>}{2\sqrt{3}}-\frac{|\downarrow\uparrow\uparrow\downarrow>}{\sqrt{3}}-\frac{|\uparrow\downarrow\downarrow\uparrow>}{\sqrt{3}}+\frac{|\uparrow\downarrow\uparrow\downarrow>}{2\sqrt{3}}+\frac{|\uparrow\uparrow\downarrow\downarrow>}{2\sqrt{3}},\\
|\rm singlet\mbox{-}2;S_{tot}^z=0>&=&\frac{|\downarrow\downarrow\uparrow\uparrow>}{2}-\frac{|\downarrow\uparrow\downarrow\uparrow>}{2}-\frac{|\uparrow\downarrow\uparrow\downarrow>}{2}+\frac{|\uparrow\uparrow\downarrow\downarrow>}{2},\\
|\rm triplet\mbox{-}1;S_{tot}^z=1>&=&\frac{|\downarrow\uparrow\uparrow\uparrow>}{2}-\frac{|\uparrow\downarrow\uparrow\uparrow>}{2}-\frac{|\uparrow\uparrow\downarrow\uparrow>}{2}+\frac{|\uparrow\uparrow\uparrow\downarrow>}{2},\\
|\rm triplet\mbox{-}2;S_{tot}^z=1>&=&\frac{|\uparrow\downarrow\uparrow\uparrow>}{\sqrt{2}}-\frac{|\uparrow\uparrow\downarrow\uparrow>}{\sqrt{2}},\\
|\rm triplet\mbox{-}3;S_{tot}^z=1>&=&\frac{|\downarrow\uparrow\uparrow\uparrow>}{\sqrt{2}}-\frac{|\uparrow\uparrow\uparrow\downarrow>}{\sqrt{2}}, \rm and\\
|\rm quintet;S_{tot}^z=1>&=&|\uparrow\uparrow\uparrow\uparrow>.\\
\label{WF}
\end{eqnarray*}
\end{widetext}

In the case of $J_2$=$J_3$=0, the ground state is singlet-1 and the first excited state is triplet-1. The second excited states are singlet-2, triplet-2, and triplet-3,  which are degenerate. The fourth excited state is the quintet. Neutron scattering observes excitations from singlet-1 to triple-t-1, triplet-2, and triplet-3 at low temperatures. With increasing temperature, excited states become populated and excitations from singlet-2 to triplets and from triplets to quintet are expected to increase.

Including non-zero $J_2$ and $J_3$ smaller than $J_1$, the ground and the first excited states remain singlet-1 and triplet-1, respectively. The first excited energy is also not affected by the finite $J_2$ and $J_3$. On the other hand, excitation energies to triplet-2 and triplet-3 become 2$J_1-J_3$ and 2$J_1-J_2$, respectively.

\begin{figure}
\includegraphics[width=8.2cm]{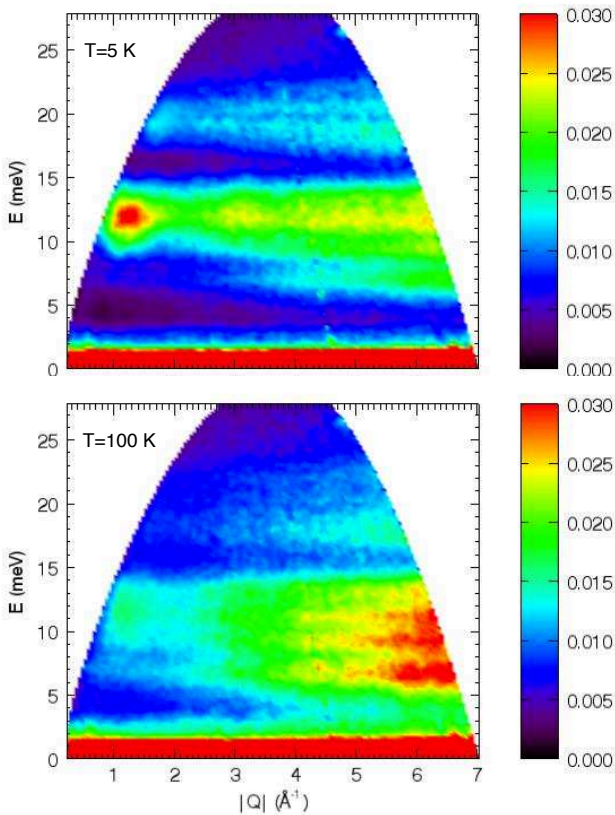}
\caption{(Color online) Color contour maps of the inelastic neutron scattering intensity $S(|Q|,E)$ for Cu$_2$PO$_4$OH powder measured with $E\rm_i$=30 meV at $T$=5 and 100 K. Background scattering measured with an empty sample can is subtracted.}
\label{E_Q_plot}
\end{figure}
\begin{figure}
\includegraphics[width=8.2cm]{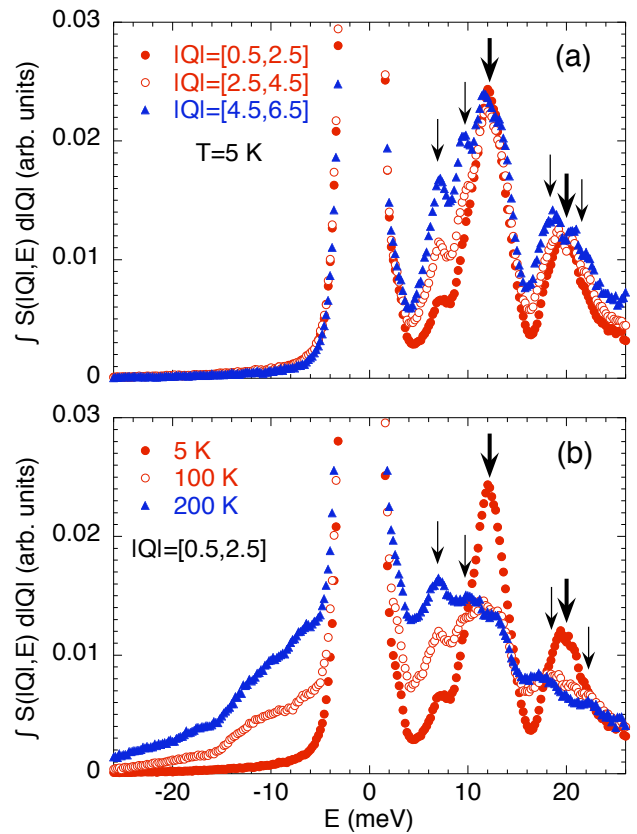}
\caption{(Color online) (a) Energy cuts through $S(|Q|,E)$ at intervals of $|Q|$=0.5$-$2.5, 2.5$-$4.5, and 4.5$-$6.5 \AA$^{-1}$ measured with $E\rm_i$=30 meV at $T$=5 K. (b) Energy cuts of $S(|Q|,E)$ integrated with respect of $|Q|$ in the range of $|Q|$=0.5$-$2.5 \AA$^{-1}$ as a function of temperature.
The integrated value is normalized by the accessible $|Q|$ range due to kinematic constraints as shown in Fig. \ref{E_Q_plot}.
The thick and thin arrows represent magnetic and phonon signals, respectively.}
\label{I_E}
\end{figure}
\section {Experimental results and Discussion}
A powder sample of undeuterated Cu$_2$PO$_4$OH with a mass of $\sim$6.5 g was prepared by the hydrothermal method as decsribed in Ref. \onlinecite{Belik07}. 
The inelastic neutron scattering experiments were carried out on a chopper neutron spectrometer ARCS, \cite{ARCS} installed at the Spallation Neutron Source (SNS) at Oak Ridge National Laboratory (ORNL). We utilized two incident energies of 30 and 60 meV. Energy resolutions at the elastic position are $\sim$1 and $\sim$ 2 meV with $E\rm_i$=30 and 60 meV, respectively. We used an annular sample cell to reduce attenuation from hydrogen and multiple scattering. The measurements were performed in a temperature range of 5 $\le T \le$ 200 K using a closed-cycle refrigerator. Measurements with an empty can were also performed to subtract the background scattering properly.

Figure \ref{E_Q_plot} shows the inelastic neutron scattering spectra $S(|Q|,E)$ from Cu$_2$PO$_4$OH powder measured with $E\rm_i$=30 meV. At $T$=5 K, two flat excitations are clearly observed at $\sim$12 and $\sim$20 meV at low $|Q|$ region below $\sim$2.5 \AA$^{-1}$. The intensities of these excitations decrease at 100 K, suggesting that these are magnetic signals. On the other hand, at $T$=100 K, the almost non-dispersive modes, which are considered to be phonons, become more distinctive around 7, 9, 11, 13, 18, and 22 meV in the high $|Q|$ region.

Figure \ref{I_E} shows several energy cuts through $S(|Q|,E)$ measured with $E\rm_i$=30 meV.
Figure \ref{I_E}(a) shows energy cuts through $S(|Q|,E)$ measured at $T$=5 K, integrated over the different $|Q|$ ranges shown in the figure.
The peaks around 7, 10, 18, and 22 meV grow with increasing $|Q|$, consistent with the behavior expected for phonons. On the other hand, the excitations at $\sim$12 and $\sim$20 meV do not change much, probably because the magnetic and phonon contributions compensate. Magnetic scattering is mostly dominant in the low $|Q|$ region and phonon scattering is superposed in the high $|Q|$ region, where magnetic intensity becomes weak.
Temperature dependence of the energy cuts integrated with respect of $|Q|$ in the range of $|Q|$=0.5$-$2.5 \AA$^{-1}$ are shown in Fig. \ref{I_E}(b). The peaks around 7, 10, and 18 meV also grow with increasing temperature, whereas the peaks around 12 and 20 meV decrease. These results confirm that the excitations around 7, 10, and 18 meV are phonons and those around 12 and 20 meV are magnetic. The peak widths of the first and the second excited magnetic state are estimated to be $\sim$3 meV, which is larger than the instrumental resolution $\Delta E(<$1 meV).
\begin{figure}
\includegraphics[width=8.2cm]{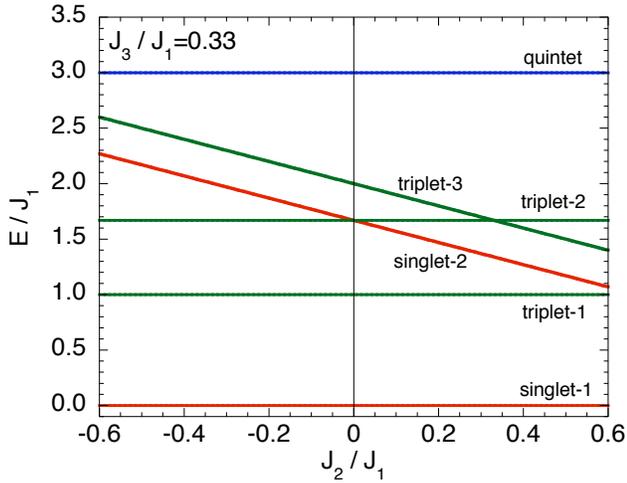}
\caption{(Color online) Excitation energies in a single tetramer as a function of $J_2/J_1$ for $J_3/J_1$ = 0.33. Two singlets, three triplets, and a quintet are shown in red, green and blue, respectively. The triplet-2 and triplet-3 become degenerate when $J_2$=$J_3$.
}
\label{diagram}
\end{figure}
\begin{figure}
\includegraphics[width=8.2cm]{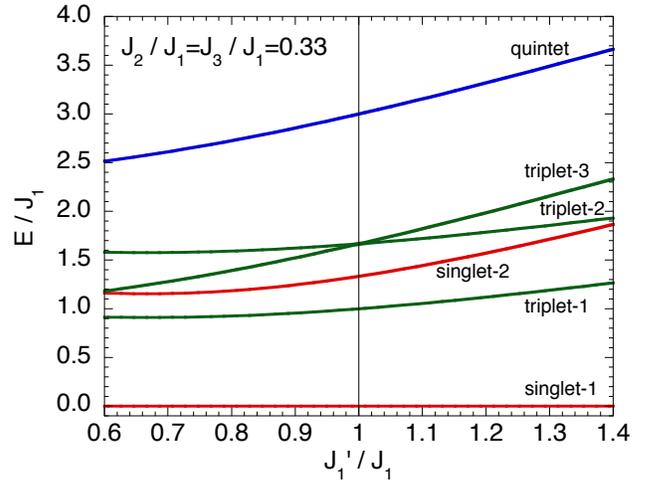}
\caption{(Color online) Excitation energies in a single tetramer as a function of $J_1'/J_1$ for $J_2/J_1$=$J_3/J_1$=0.33. Two singlets, three triplets, and a quintet are shown in red, green and blue, respectively.
}
\label{J1prime}
\end{figure}

As described in Sec. II, two magnetic excitations are expected at $E_1=J_1$($\sim$12 meV) and $E_2=2J_1$($\sim$24 meV) in the case of $J_2$=$J_3$=0. Since our results show that the ratio $|E_2/E_1|\sim$5/3, additional intratetramer interactions should be included. We first consider the simpler model with $J_1$=$J_1'$. In this case, the ratio of $\sim$5/3 is realized when $J_2$ or $J_3$ becomes $\sim$0.33$J_1$. Furthermore, since only two excitations are observed, the second and the third excited states are considered to be approximately degenerate. In Fig. \ref{diagram}, the excitation energies for a fixed value of $J_3/J_1$=0.33 are shown as a function of $J_2$. The two triplet states are degenerate only when $J_2$=$J_3$. Our results indicate that $E_1=J_1\sim$12 meV and $E_2$=2$J_1-J_3$=2$J_1-J_2$$\sim$20 meV. Therefore, it is calculated that $J_2$=$J_3\sim$ 4 meV, suggesting that both diagonal couplings are competing with the main interaction $J_1$. The finite antiferromagnetic coupling of $J_2$ is reasonable, because $J_2$ is mediated by the Cu-O-Cu superexchange coupling with a bond angle of 97$^\circ$. However, it is unexpected that $J_3$, which is mediated by the Cu-O-O-Cu super-superexchange interaction, is relatively large.

As described in Sec. I, the structural distortion at 160 K, which makes $J_1$ and $J_1'$ inequivalent, also changes the energy levels.
We performed further analytical calculations including $J_1'$. The energy levels of a single tetramer for $J_1'/J_1=r$ and $J_2/J_1$=$J_3/J_1$=0.33 are obtained as follows:
\begin{eqnarray*}
E_{\rm singlet\mbox{-}1}&=&\frac{1}{6}J_1(-4-3r-2\sqrt{7-12r+9r^2}),\\
E_{\rm singlet\mbox{-}2}&=&\frac{1}{6}J_1(-4-3r+2\sqrt{7-12r+9r^2}),\\
E_{\rm triplet\mbox{-}1}&=&\frac{1}{6}J_1(-2-3r),\\
E_{\rm triplet\mbox{-}2}&=&\frac{1}{6}J_1(2-3r),\\
E_{\rm triplet\mbox{-}3}&=&\frac{1}{6}J_1(-4+3r), \rm and\\
E_{\rm quintet}&=&\frac{1}{6}J_1(4+3r).
\label{EL2}
\end{eqnarray*}
Figure \ref{J1prime} shows the excitation energies for $J_2/J_1$=$J_3/J_1$=0.33 as a function of $J_1'/J_1$. The excitation energies from singlet-1 to triplet-2 and -3 split as $J_1'/J_1$ deviates from 1, causing the second excited state to split or broadened without changing the width of the first excited state. 
Analytical calculations in a first order approximation show that the bandwidths of the triplet-1, triplet-2, and triplet-3 are 2/3($J_4$+$J_5$), $J_5$/3, and $J_4$/3, respectively. Therefore, the bandwidths of the first and the second excited state are 2/3($J_4$+$J_5$) and a value less than ($J_4$+$J_5$)/3, respectively. This means that the peak width of the first excited state should be at least twice as large as that of the second excited state.
The experimental results show that the energy width of the second excitation peak is similar to that of the first excited state. This indicates that $J_1'$ can be only slightly larger or smaller than $J_1$.

$J_1'$, $J_4$, and $J_5$ are estimated from the observed excitation widths as follows. It is noted that a super-superexchange coupling (bond distance: 5.4 \AA) was assumed as an intertetramer coupling between Cu2. \cite{Belik07} Here, we assume that $J_5$ is a superexchange coupling between Cu2 with a bond distance of 3.06 \AA\ and a Cu-O-Cu bond angle of 100$^\circ$. This interaction is considered to be antiferromagnetic because of the bond angle.
The excitation widths are observed to be $\sim$3 meV, which is much broader than the instrumental resolution ($<$1 meV). This suggests that the intrinsic excitation band width is about 2.8 meV. For the first excited state, the band width 2/3($J_4$+$J_5$) should correspond to the excitation width 2.8 meV. Therefore, $J_4$+$J_5$$\sim$4.2 meV.
 Considering that $J_2$ has been determined to be nonnegligible in this study, it would be reasonable to assume that $J_4$ is also finite and similar to $J_2$, since the Cu-O-Cu superexchange paths of $J_2$ and $J_4$ should be similar. If this is the case, $J_4$ is assumed to be $\sim$4 meV and $J_5$ to be very small. Then, the second excited state should have the width of at least $\sim$1.3 meV. An additional broadening probably originate from the peak splitting due to $J_1'$. In the case of $J_1'$/$J_1$=0.9 and 1.1, the peak broadening is estimated to be about 3 meV, including the instrumental resolution. Therefore, the difference between $J_1$ and $J_1'$ is about 10\%. Here, we define that $J_1>J_1'$. The interactions with $J_1\sim$12.5 meV, $J_1'\sim$11.3 meV and $J_2\sim J_3\sim J_4\sim$4 meV reasonably reproduce the two broad magnetic excitations observed.

\begin{figure}
\includegraphics[width=8.2cm]{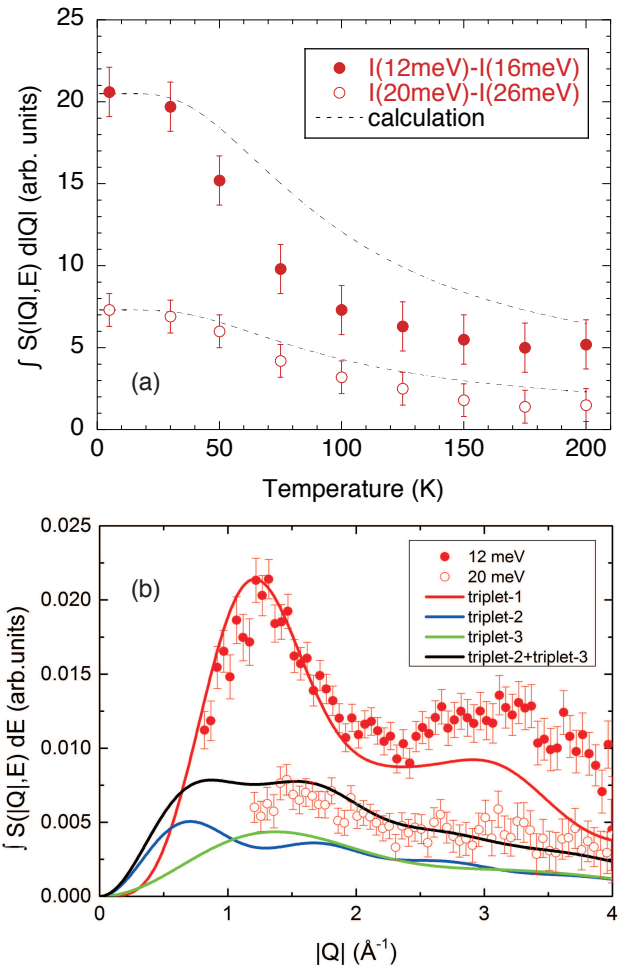}
\caption{(Color online) (a) Temperature dependences of the magnetic excitation intensities at 12 and 20 meV measured with  $E\rm_i$=30 meV. The intensities are integrated with respect to $|Q|$ in the range of 1.0$\le|Q|\le$ 2.5 \AA$^{-1}$ at 12 meV and 1.7$\le|Q|\le$ 2.5 \AA$^{-1}$ at 20 meV. Background intensities at 16 and 26 meV are subtracted from the 12 meV and 20 meV data, respectively. The broken lines are the calculated values using Eq. (1). (b) $S(|Q|,E)$ integrated with respect to energies $E$=11.5$-$12.5 and 19.5$-$20.5 meV measured with $E\rm_i$=60 meV. The intensity at $E$=4.5$-$5.5 meV, where phonon scattering is not very intense, is subtracted as a constant background. Therefore, the data contain phonon scattering especially at high $|Q|$.}
\label{I_TQ}
\end{figure}
The temperature dependences of the scattering intensities between the singlet ground state and the first (12 meV) and second excited states (20 meV) are plotted in Fig. \ref{I_TQ}(a). Background intensities at 16 and 26 meV were subtracted from the excitation data at 12 and 20 meV, respectively.
The scattering intensity is proportional to the thermal population factor $R_{m}(T)$ for the initial state with $E_m$, given by

\begin{eqnarray}
R_{m}(T)=\frac{{\rm exp}(-E_m/k_BT)}{\displaystyle\sum_{i=1} {\rm exp}(-E_i/k_BT)} .
\label{RT}
\end{eqnarray}
For the 12 meV and 20 meV excitations, the singlet-1 ground state is considered as the initial state. In Fig. \ref{I_TQ}(a), the observed and calculated intensities are normalized at 5 K. Equation 1 gives higher intensities than the excitation intensities observed. This is probably because the tetramers are not isolated well, consistent with the finite intertetramer interactions as discussed above.

Magnetic excitations between excited states are also expected with increasing temperature, as described in Sec. II. Excitations from singlet-2 to triplet-2 and -3, from triplet-1 to quintet, and from triplet-2 and -3 to quintet are expected with excitation energies of $\sim$0.4$J_1$($\sim$5 meV), $\sim$2.0$J_1$($\sim$25 meV), and $\sim$1.3$J_1$($\sim$16 meV), respectively. As shown in Fig. \ref{I_E}(b), intensities at $\sim$5 and $\sim$16 meV increase with increasing temperature. On the other hand, the intensity at $\sim$25 meV does not change much with temperature. Since tails of the phonon peaks are also superposed at these energies, it is difficult to discuss the temperature dependence of these intensities quantitatively.

The $|Q|$ dependences of the excitation intensities between the singlet ground state and the first and second excited states are plotted in Fig. \ref{I_TQ}(b). The two excitation states (triplet-2 and -3) are nearly degenerate at 20 meV.
The calculated intensity corresponds to the square of the powder averaged tetramer structure factor multiplied by the magnetic form factor of Cu$^{2+}$ spins. The calculated intensities reproduce the low-$|Q|$ data ($\le$$\sim$2.4 \AA$^{-1}$) reasonably well. Phonon scattering overlaps at higher $|Q|$. However, the broad peak position at $\sim$3 \AA$^{-1}$ of the 12 meV excitation is well reproduced, since the phonon intensity just gradually increases with increasing $|Q|$. These results confirm that the first and second excited states are triplet-1 and a nearly degenerate state of triplet-2 and -3, respectively, suggesting that the interactions of $J_1\sim$12.5 meV, $J_1'\sim$11.3 meV and $J_2\sim J_3\sim J_4\sim$4 meV are accurate.

\section {Summary}
Inelastic neutron scattering experiments have been carried out on Cu$_2$PO$_4$OH, consisting of diamond-shaped isotropic quantum spin tetramers. We have found two nearly dispersionless magnetic excitations at $\sim$12 and $\sim$20 meV with broad excitation width. It was estimated that the coupling constants are $J_1\sim$12.5 meV, $J_1'\sim$11.3 meV, and $J_2\sim J_3\sim J_4\sim$4 meV. Thus, the $S$=1/2 diamond-shaped antiferromagnetic tetramer consists of two inequivalent main interactions and finite competing diagonal couplings. Compared to the other tetramer materials described in Sec. I, Cu$_2$PO$_4$OH is unique in a sense that the diamond-shaped tetramers are connected quasi-one dimensionally.

\begin{acknowledgments}
The research at ORNL's SNS was sponsored by the Scientific User Facilities Division, Office of Basic Energy Sciences, U. S. Department of Energy.
\end{acknowledgments}


\begin{thebibliography}{}
\bibitem{Haral05} J. T. Haraldsen, T. Barnes, and J. L. Musfeldt, Phys. Rev. B \textbf{71}, 064403 (2005).
\bibitem{Furrer13} A. Furrer and O. Waldmann, Rev. Mod. Phys. \textbf{85}, 367 (2013).
\bibitem{ethredge95}K. M. S. Etheredge and S. -J. Hwu, Inorg. Chem. \textbf{34}, 5013 (1995).
\bibitem{ulutagay03}M. Ulutagay-Katin, S. -J. Hwu, and I. A. Clayhold, Inorg. Chem. \textbf{42}, 2405 (2003).
\bibitem{Belik05}A. A. Belik, M. Azuma, A. Matsuo, M.-H. Whangbo, H.-J. Koo, J. Kikuchi, T. Kaji, S. Okubo, H. Ohta, K. Kindo, and M. Takano, Inorg. Chem. \textbf{44}, 6632 (2005).
\bibitem{Belik06}A. A. Belik, M. Azuma, A. Matsuo, T. Kaji, S. Okubo, H. Ohta, K. Kindo, and M. Takano, Phys. Rev. B \textbf{73}, 024429 (2006).
\bibitem{hase97}M. Hase, K. M. S. Etheredge, S. J. Hwu, K. Hirota, and G. Shirane, Phys. Rev. B \textbf{56}, 3231 (1997).
\bibitem{Muller00} A. M\"{u}ller, Ch. Beugholt, P. K\"{o}gerler, H. B\"{o}gge, S. Budko, and M. Luban, Inorg. Chem. \textbf{39}, 5176 (2000).
\bibitem{Furrer10} A. Furrer, K. W. Kr\"{a}mer, Th. Str\"{a}ssle, D. Biner, J. Hauser, and H. U. G\"{u}del, Phys. Rev. B \textbf{81}, 214437 (2010).
\bibitem{Baster02} R. Baster, G. Chaboussnt, A. Sieber, H. Andres, M. Murrle, P. K\"{o}gerler, H. B\"{o}gge, D. C. Crans, E. Krickemeyer, S. Janssen, H. Mutka, A. M\"{u}ller, and H. U. G\"{u}del, Inorg. Chem. \textbf{41}, 5675 (2002).
\bibitem{Belik07}A. A. Belik, H.-J. Koo, M.-H. Whangbo, N. Tsujii, P. Naumov, and E. Takayama-Muromachi, Inorg. Chem. \textbf{46}, 8684 (2007).
\bibitem{kuo08} C. N. Kuo and C. S. Lue, Phys. Rev. B \textbf{78}, 212407 (2008).
\bibitem{Belik11}A. A. Belik, P. Naumov, J. Kim, and S. Tsuda, J. Solid State Chem. \textbf{184}, 3128 (2011).
\bibitem{ARCS}D. L. Abernathy, M. B. Stone, M. J. Loguillo, M. S. Lucas, O. Delaire, X. Tang, J. Y. Y. Lin, and B. Fultz, Rev. Sci. Instrum. \textbf{83}, 15114 (2012).
\end{thebibliography}
\end{document}